\documentstyle[12pt,epsf]{article}
\begin{document}

\begin{flushright}
PSI-PR-95-20\\
\end{flushright}

\begin{center}

{\bf \large
On the Relationship of the Scaled
Phase Space and Skyrme-Coherent State Treatments of
Proton Antiproton Annihilation at Rest }
\end{center}
\normalsize
\medskip

\begin{center}
R. D. Amado,$^{a,b}$ F. Cannata,$^{b,c}$ J-P. Dedonder,$^{b,d}$
M. P. Locher,$^{b}$ \\  Yang Lu,$^a$ and V. E. Markushin $^b$\\
\end{center}
\small
$^a$ {\it Department of Physics, University of Pennsylvania,
 Philadelphia, PA 19104, USA      \\}
$^b$ {\it Paul Scherrer Institute, CH-5232 Villigen PSI, Switzerland\\}
$^c$ {\it Dipartimento di Fisica and INFN,
  I-40126 Bologna, Italy \\}
$^d$ {\it Laboratoire de Physique Nucl\'{e}aire, Universit\'{e} Paris 7 - 
Denis Diderot, F-75251 Paris Cedex 05 and Division de 
Physique Th\'{e}orique, IPN
F-91406 Orsay, France \footnote{The Division de
Physique Th\'{e}orique is a Research Unit of the
Universities of Paris 11 and 6 associated to CNRS.
 }}\\
\normalsize

\centerline{{\bf Abstract}}
\noindent
We discuss pion multiplicities and single pion momentum
spectra from proton antiproton annihilation at rest.
Both the scaled phase space model and the Skyrme-coherent
state approach describe these observables well. In the 
coherent state approach the puzzling size of the scale   
parameter relating the phase space integrals for different
multiplicities is replaced by a well defined weight
function.  The strength of this function is determined
by the intensity of the classical pion field  
and its spatial extent is of order 1 fm.

\noindent
\newpage

The phenomenology of proton antiproton annihilation at rest is a 
venerable subject going back to Fermi \cite{Fermi}. The connection
of successful phenomenology with  dynamical theory is difficult
because low energy annihilation into
pions is squarely in the domain of nonperturbative QCD.  Recently
we have shown that Skyrmion dynamics, a model based on a classical
approach to QCD, combined with coherent states to account
for the quantum nature of the fields, gives a remarkably good
account of many of the features of low energy annihilation and
does so with essentially no free parameters \cite{us}.  In this
note we compare the Skyrme-coherent state approach with the
scaled form of the statistical phase space method
introduced by Fermi. In particular we  
explore what parts of these  approaches 
are responsible for the general agreement, what parts are
needed for the details, and what parts serve to bridge the
gap from phenomenology to dynamical model.      

\section{Scaled Phase Space}

The most important constraint on any calculation of 
pions from proton-antiproton annihilation at rest 
is that of energy momentum conservation.  That is, 
the rate for finding $n$ pions of three-momenta
$\mbox{\bf k}_i$, with $i=1...n$, should be proportional
to the differential 
phase space factor $\rho_n(s, \{\mbox{\bf k}_i\})$ given
by
\begin{equation}
\rho_n(s,\{\mbox{\bf k}_i\}) = \delta^4(k_t -\sum_{i=1}^n k_i)
\prod_{i=2}^n \frac{d^3\mbox{\bf k}_i}{2 \omega_i}
\end{equation}
where $s = (k_t)^2$, $k_t$ is the total four-momentum of the
annihilating pair and $\omega_i$ is the energy of the
$i$-th pion, $\omega_i= \sqrt{\mbox{\bf k}_i^2+\mu^2}$, with
$\mu$ the pion mass.  For annihilation at rest 
$k_t = (2M, \mbox{\bf 0})$, with $M$ the nucleon
mass. The ``phase space only" (PSO) assumption is that 
all other aspects of the annihilation process depend
very weakly on $\{\mbox{\bf k}_i\}$ so that the entire dependence
is given by $\rho_n(s,\{\mbox{\bf k}_i\})$.  For example
the single pion momentum spectrum for
$n$ pions is obtained by integrating $\rho_n(s,\{\mbox{\bf k}_i\})$
over all but one of the final momenta.  We will return
below to the result of that integration.  
 
If we want to compare the branching ratio for
$n$ to that for  $n+1$ pions we observe that the corresponding integrals
in (1) differ in dimension by two units of momentum. Thus to compare them,
all other things being equal, we must construct a quantity of uniform
dimension, $R(n)$, by scaling the total phase space for $n$ pions 
we write
\begin{equation}
R(n) = \frac{(L)^{2n}}{n!} \int \rho_n(s,\{\mbox{\bf k}_i\})
\end{equation}
where $L$ is a length
and an $n$ independent overall normalization factor
is set equal to 1. We call this picture for
 relating multiplicities by a dimensional scaling
``scaled phase space,"  SPS. In terms of $R(n)$ one can calculate
the average number of pions in the SPS picture by
\begin{equation}
\hat{n} = \frac{\sum_n n R(n)}{\sum_n R(n)}
\end{equation}
This $\hat{n}$ depends on $L$ which can be varied to give
the experimental value,  $\hat{n} = 5$.  With the
scaling length  fixed by
this empirical constraint, the resulting pion multiplicity distribution,
the normalized $R(n)$, looks very much like the experimental
one. Fig.~1 shows the pion multiplicity distribution
calculated in the SPS 
formalism\footnote{The non-relativistic form
of phase space leads to very
similar distributions, as we have checked and as 
was already reported in \cite{Dover}  }
with $L = 1.2$ fm,
adjusted to give the correct average pion number.
It has a gaussian shape
with  an average of 5 (put in by hand)
and a variance of 1.  Note no statistical assumption has been
made to obtain the observed gaussian distribution
or the correct variance. Fig.~1 also displays   
the corresponding Skyrme-coherent
state calculation of Sect.~2 and the measured distribution. Both
calculations agree excellently with the data.

\begin{figure}[htbp]
\begin{center}
\vspace*{-10mm} 
 \mbox{\epsfysize=6cm\epsffile{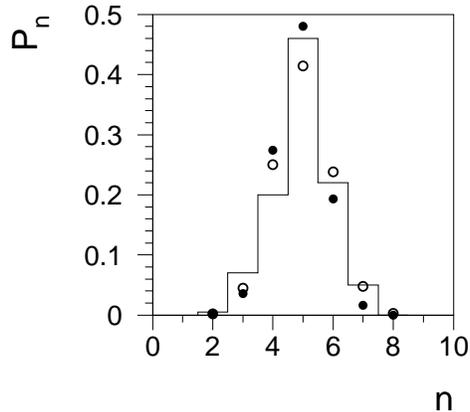}}
\vspace*{5mm} 
\end{center}
\caption{\label{Fig1}%
Pion multiplicity distribution in proton antiproton
annihilation at rest. The open dots are from the scaled phase space
model and the solid dots the from Skyrme-coherent state approach.  
Both have been fixed to give an average pion multiplicity of 5.  The
histogram is the data presented in \protect\cite{Dover,Sedlak}.
}
\end{figure}

 In earlier treatments of the scaling model $L$ was connected to 
a volume \cite{Dover,old} by  
$V=(2 \pi L)^3$.  The $2 \pi$ comes from the normalization
of phase space density.  
The volume needed to give $\hat{n}=5$ turns out to be very large
by nucleon standards, of order $(2\pi)^3$ fm$^3$.  There 
is much discussion in the early literature about the meaning of
this large volume \cite{Dover,old}. 
An alternate picture is that one should write 
$V = g^2 \cdot v$ 
where $v$ is a reasonable volume ($\sim \frac{4 \pi}{3}\mu^{-3}$
= $12.5$ fm$^3$) while $g$
is a dimensionless number that gives the amplitude for emitting
one more pion. Now it is $g$ that is large. The notion that
the amplitude for emitting $n$ pions should be proportional to
$g^n$, the coupling constant to the $n$-th power, makes sense
perturbatively, but is being employed here for $g$ large.
Annihilation is nonperturbative. Thus the
dynamical origin of the SPS picture remains obscure, even though
it can certainly fit the pion multiplicity distribution with one
free parameter.  

 Historically statistical approaches were
unable to account for the pion multiplicity distribution
in a pions only picture 
with a ``reasonable"  scaling parameter, or to connect the
picture to more fundamental theory.  Adding heavy mesons that
decay sequentially to pions helps the phenomenology, but
at the expense of more free parameters \cite{van}. There have also
been attempts to make thermodynamic models of annihilation without
imposing the constraints of energy-momentum conservation \cite{Blumel}.
These do poorly for the pion multiplicity distribution even when
$\hat{n}$ is fixed at 5. This serves to further emphasize
the primary importance of the four-momentum phase space constraint.  

For fixed multiplicity, $n$, the PSO picture has no dynamics.  
Relating the amplitude for
one $n$ to the next implies dynamical
assumptions outside PSO as in the 
scaled phase space scheme SPS, discussed above.
For a natural candidate that introduces genuine dynamics we turn to
the Skyrme-coherent state approach. It relates the weights and momentum
spectra of $n$ pion emission in a transparent way to the properties
of the underlying annihilation process.

\section{Skyrme-Coherent State Approach}

In the Skyrme-coherent state approach one uses a classical meson
field theory in which baryons appear as topologically stable
solitons to model the dynamics of annihilation.  This picture, invented
by T.H.R. Skyrme \cite{Skyrme}, is connected to QCD in the limit
of a large number of colors, and through that to the long wave
length or nonperturbative limit of QCD \cite{tHooft,Witten}.
It is found that in the Skyrme approach, annihilation
proceeds very rapidly leading to a burst of 
relatively intense classical pion radiation \cite{SSLK,SWA}. 
 To connect with the physical pion quanta of
experiment, that classical wave is used to generate a quantum mechanical
coherent state \cite{Klauder}. There are two steps to this
process, the use of Skyrme dynamics to generate
the classical pion radiation from annihilation, and the subsequent
quantization of that radiation using coherent states.  The
circumstances of annihilation seem particularly well suited to 
this combination.  

 A standard coherent state does not have fixed
four-momentum, but as we saw above, that constraint is crucial.  Hence
the coherent state must be projected onto a state of definite 
energy-momentum \cite{HornSilver}, and if we are interested in
pion charge ratios, a state of definite isospin \cite{UCSB}.
We have developed the formalism for doing all this \cite{us}.
A pion coherent state contains an exponential in
the pion field creation operator. It is a single
quantum state containing all pion numbers.  This is the
physics appeal of the coherent state approach, namely
that all the pion channels are collected into a single state.
Thus questions about the relation of the rate or spectrum in
the $n$ pion channel to that in the $n+1$ channel, are naturally
answered.  No new parameters or assumptions are needed to address
them.  Since the coherent state approach can also be generalized to
include energy-momentum conservation, the results discussed in
Sect.~1 come out, but now with a clear origin for
the relationship among the channels.    

In the Skyrme-coherent state approach,
the difficult dynamics of nucleon-antinucleon 
interaction and subsequent annihilation into pions is done 
classically, and quantum mechanics only enters to describe the
propagation of the coherent state after the fields have 
reached the radiation zone, where they are non-interacting. Although
this program is far simpler than the corresponding full quantum
QCD calculation of annihilation, it is still complicated
to execute with the
classical, nonlinear field equations of Skyrme and 
up to now has not been fully implemented.      
We have not studied the development of the annihilation
process itself, but rather have begun with an assumed
initial spherically symmetric 
configuration of classical pion field.    It is
in this initial pion configuration, that free parameters
 enter.  The remaining dynamics is completely determined.
We take a simple initial pion configuration
characterized by a size and magnitude.  The magnitude 
is fixed by the total energy of the
system, $2M$, leaving only the size to be determined.
This is fit to the average pion multiplicity, or
equivalently, the inclusive single pion momentum distribution.                
We find a size of order 1~fm, a completely reasonable result.
Note that if at some future date we are able to do the
Skyrme calculation of annihilation from the beginning, there
would be no free parameters whatsoever.  

The introduction of a finite source size for the pion radiation
leads to a form factor for pion emission $f(k_i)$.  This
form factor is the Fourier transform of the classical asymptotic
pion field.  As such it is similar to, but not identical to the Fourier
transform of the pion
source distribution because of 
the intervening Skyrmion dynamics. 
For the coherent state formalism of \cite{us} we must replace (2) by
\begin{equation}
R^{COH}(n) =  \frac{1}{n!} \int \prod_{i=2}^n |f(k_i)|^2 
\rho_n(s,\{\mbox{\bf k}_i\})
\end{equation}
(without the complication of isospin). This expression is
completely specified by the coherent state formalism through
the form factors. Hence in
the Skyrme-coherent state approach the ratios of multiplicities
are controlled not by some arbitrary volume, but by the 
intensity of the classical radiation field.   
In this work we take the analytic form for the form factor
which we have used before \cite{us}.
Adapted to the relativistic phase space convention it reads
\begin{equation}
|f(k)|^2 = \frac {2 C_0 \mbox{\bf k}^2}{
\omega (\mbox{\bf k}^2+\alpha^2)^2 (\omega^2+\alpha^2)^2}
\label{fk}
\end{equation} 
with $\alpha = 3.0 \mu$ 
and $C_0= 0.061$ GeV$^5$. $C_0$ has been determined by requiring that 
the energy release of the classical pion source is $2 M$.
The size parameter $\alpha$ has been adjusted to the measured average 
multiplicity $\hat{n}=5$ imposing energy 
momentum-conservation\footnote{\label{a}%
\mbox{Without energy momentum conservation the classical mean pion 
multiplicity} 
\mbox{$\int |f(k)|^2 d^3\mbox{\bf k} (2 \omega)^{-1}$} is $4.5$ for $\alpha = 3 \mu$.}
as in \cite{us}.

Note that any classical field theory capable of describing
the evolution of classical pion radiation from 
annihilation will lead to
a quantum coherent state and to form factors.  We 
emphasize the Skyrme method because 
it is the only one we know that naturally
gives annihilation and subsequent pion radiation.   

For the pion multiplicity
spectrum we have seen in Fig.~1 that the introduction of form factors
in the coherent state approach leads to results equivalent to those of 
the SPS model and in agreement with experiment. However now the
relative weighting of different multiplicities has a simple physical
origin in the strength and size of the initial pion source. The strength
is determined by the energy release and the spatial distribution 
corresponding to (\ref{fk}) has 
an r.m.s. of 0.7~fm which is reasonable. 
Next we calculate the single pion momentum spectra for the pion
multiplicities individually, by integrating the phase space
over all but one of the pion momenta 
\begin{eqnarray}
 \frac{dN_n(K)}{dK} & = & \frac {1}{n!} 
                        \int \delta(K - |K_1|) \prod_{i=2}^n 
                        |f(k_i)|^2 \rho_n(s,\{\mbox{\bf k}_i\}) \\
  \frac{dN(K)}{dK} & = &  \sum_n \frac{dN_n(K)}{dK}  \nonumber
\end{eqnarray}
where we use the notation $K = |\mbox{\bf k}| $.

\begin{figure}[htbp]
\begin{center}
\vspace*{-12mm} 
 \mbox{\epsfysize=12cm\epsffile{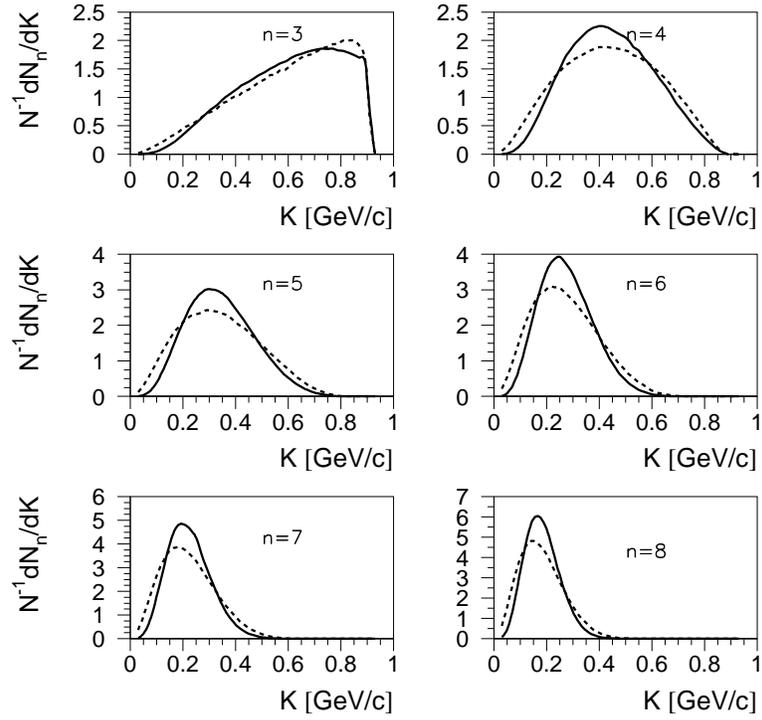}}
\vspace*{-18mm} 
\end{center}
\caption{\label{Fig2}%
The single pion inclusive momentum distribution ($dN_n/dK$) for 
annihilation to channels with  pion multiplicity $n=3$ to $8$. The
solid line is from the Skyrme-coherent state approach and 
the dotted line is phase space only. All distributions are
normalized to one.
}
\end{figure}

\begin{figure}[htbp]
\begin{center}
 \mbox{\epsfysize=8cm\epsffile{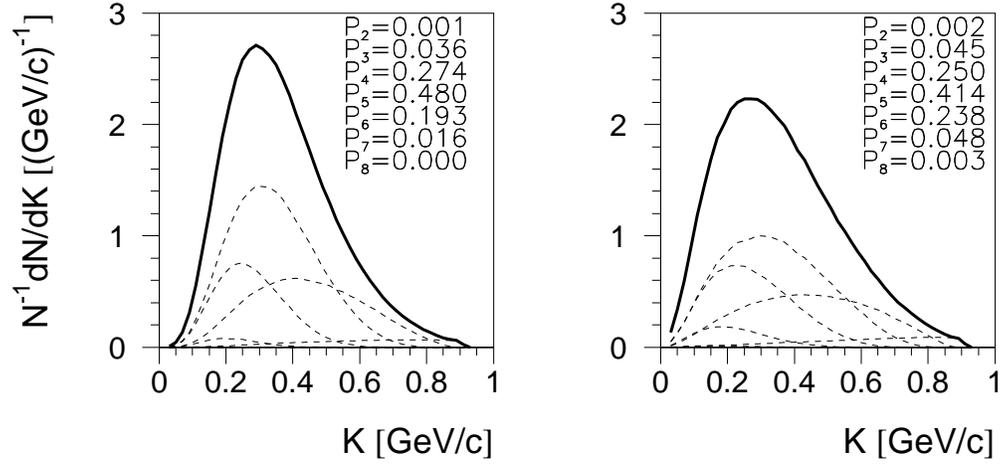}}
\vspace*{-12mm} 
\end{center}
\caption{\label{Fig3}%
The single pion inclusive momentum distributions
$dN_n/dK$ for each multiplicity weighted by the probability
of that multiplicity and the sum of these, $dN/dK$, which is the
full inclusive pion momentum spectrum.
The Skyrme-coherent state case  is shown on the left
and the scaled phase space case on the right. The 
probabilities of the different multiplicities are shown on
the graph. The summed spectrum is normalized to one.
}
\end{figure}

\begin{figure}[htbp]
\begin{center}
 \mbox{\epsfysize=8cm\epsffile{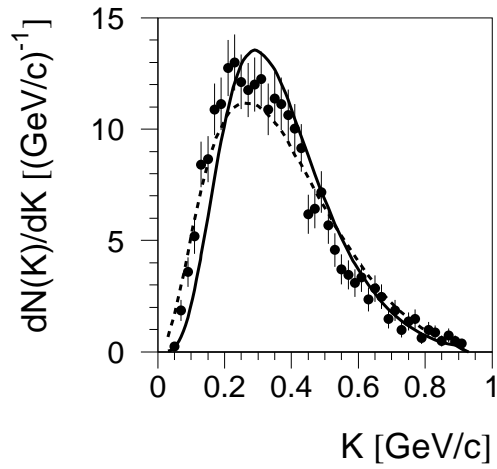}}
\vspace*{-12mm} 
\end{center}
\caption{\label{Fig4}%
The single pion momentum distribution, $dN/dK$,
summed over all multiplicities, normalized to the
total number of pions.  The dotted line is from the scaled
phase space model, the solid line from the Skyrme-coherent
state approach and the data are from \protect\cite{Sedlak}. 
}
\end{figure}

In Fig.~2 we show the single particle
spectra ($dN_n/dK$) for $n=3$ to $8$ final pions for 
the scaled phase space 
model SPS and for the coherent state approach \cite{us,LuAmado}.
For each $n$, each graph is normalized to 1.    We see that there is some
difference in detail between the SPS and the coherent state
approach, but they are quite similar in general shape.  
We have been unable to find recent data with which to
compare these pion spectra for fixed $n$.
In Fig.~3 we show the single pion
momentum spectra of Fig.~2 with their 
correct relative weights.  The probabilities of the 
different multiplicities, $P_n$ are listed on the figure.
For the coherent state case the relative weights
come out of the dynamics, for the phase space case they
are put in through the scaling, fit to give $\hat{n} =5$.
We see that in both cases only the $n=4,5,6$ multiplicities
have substantial weight. Also shown in Fig.~3 are the 
weighted sum (normalized to one)
of the momentum distributions in each 
multiplicity, the inclusive single pion momentum
spectrum.  In Fig.~4 we show that inclusive pion spectrum
again (this time normalized to the total number of direct pions)
comparing the SPS result, the coherent state result
and the data reported in \cite{Dover}.   
The two calculations agree qualitatively,
giving an equivalently good
account of the data.

\section{Conclusions}

We have seen that the principal features of the pion multiplicity 
distribution and of the pion momentum spectrum in proton
antiproton annihilation at rest can come from phase space so long as one
connects the probabilities for different pion multiplicities.  
This can be done in an ad hoc way in 
the scaled phase space picture by introducing 
a scaling volume or in a dynamically
motivated  way in the context of the Skyrme-coherent
state approach.  The scaling volume needed in the phase space picture, 
$(2 \pi)^3$~fm$^3$, is an order of magnitude too large. 
No such volume interpretation is 
 required in the Skyrme approach. Rather a form factor appears naturally
the strength of which is fixed by the magnitude of the 
classical pion field, or equivalently by the energy released
in annihilation, $2M$, and the range of which is fit to
get an average pion number of 5, yielding a size of about $0.7$ fm. 
Furthermore
if a complete calculation of annihilation using Skyrme 
dynamics  were carried out (a difficult but not impossible
task) there would be no free parameters in its description of annihilation.

Finally we should point out that we have only discussed 
inclusive pion multiplicity and
momentum spectra here.  The Skyrme-coherent state picture has also been used
to calculated pion charge branching ratios  and extended to include
vector mesons all with no new free parameters \cite{LuAmado}. 
These vector mesons
are generated by the extended Skyrme dynamics since we take the
initial configuration to be pions only.  The calculated charge
branching ratios and branching ratios into vector mesons ($\rho$ and
$\omega$) are in qualitative agreement with experiment \cite{Dover,Sedlak}. 
A corresponding 
 phase space only calculation would require additional free parameters to
 generate these branching ratios. Further afield, two pion correlations,
which have been 
discussed in the Skyrme-coherent state picture \cite{ACDLL94,LA2},
find no natural explanation in the scaled phase space approach.  

\section{Acknowledgments}   

RDA, FC, and J-PD thank the theory group of the Division of Nuclear
and Particle Physics of the Paul Scherrer Institute for, once again,
providing a stimulating environment for this work.
The work of RDA is partially
supported by the United States National Science Foundation.

\end{document}